\documentclass[twocolumn,showpacs,preprintnumbers,unsortedaddress]{revtex4}

\usepackage{amsmath,amssymb}
\usepackage{graphicx}
\usepackage{dcolumn}
\usepackage{bm}

\begin{document}

\title{Qubit transfer between photons at telecom and visible wavelengths in a slow-light atomic medium}

\author{A.Gogyan}
\email{anahit.gogyan@u-bourgogne.fr}
\affiliation{Institute for Physical Research, Armenian National Academy of Sciences, Ashtarak-2,
0203, Armenia}
\affiliation{Institut Carnot de Bourgogne, UMR CNRS 5209, BP 47870, 21078 Dijon, France}

\begin{abstract}
We propose a method that enables efficient conversion of quantum information frequency between different regions of spectrum of light  based on recently demonstrated strong parametric coupling between two narrow-band single-photon pulses propagating in a slow-light atomic medium \cite{sis}. We show that an input qubit at telecom wavelength is transformed into another at visible domain in a lossless and shape-conserving manner while keeping the initial quantum coherence and entanglement. These transformations can be realized with a quantum efficiency close to its maximum value.
\end{abstract}

 \maketitle

Photons in the telecommunication band are ideally suited for carrying quantum information over large-scale quantum network. They interact weakly with surrounding environment displaying low decoherence while propagating  through telecom fibers, which support the light propagation minimal losses at a wavelength around $1.5 \mu m$. At the same time, none of the atomic based quantum memories that have been demonstrated so far operate at this wavelength, thus impeding the interfacing of quantum communication lines with photon-memory units which is required for transferring the photonic qubits into atomic qubits and vise versa. One possible way to overcome this problem is to use wavelength conversion techniques, which have been hitherto realized both in nonlinear crystals  via parametric up-conversion with preserving a quantum state {\cite{huang, giorgi, kwiat, albota, tanzilli}} and in the atomic ensembles, where a technique for light storage and its subsequent retrieval \cite{fleis} at another optical frequency under the conditions of electromagnetically induced transparency (EIT) \cite{harris, mfleis} was employed \cite{zibrov,wang}. However, both methods are confronted with severe challenges. In the former case, the main limitation is that the light emitted in the crystals has too broad linewidth ($\sim$ 10nm) and low spectral brightness to be able to effectively excite atomic species, while in the latter case of atomic  experiments, to date no true demonstration of information-preserving frequency conversion has been given. The major difficulties inherent to atomic schemes is that the purity of the stored photon is hardly preserved, that leads to unavoidable losses and shape distortion of the quantum light pulse during its storage and retrieval by means of EIT.

Recently we have proposed a new method for quantum frequency conversion (QFC) \cite{gogmal} free from the above drawbacks. Our method makes use of recently demonstrated \cite{sis} efficient parametric coupling between two single-photon pulses with small frequency difference propagating in a slow-light medium of three-level atoms at different group velocities.  However, this method is not applicable for qubit transfer between IR and visible photons, since large frequency difference entails large difference between group velocities of the weak pulses that restricts their interaction time in the medium and, hence, strongly reduces the probability to successfully transfer the qubits. Note, that the same problem with the large frequency difference arises also in the method recently proposed in \cite{lvovsky}. In this paper we generalize our scheme where the required frequency conversion is easily realized.

The essence of our method is the following. An ensemble of cold atoms interacts with two quantum fields on the transitions $3 \rightarrow 2$ and $0 \rightarrow 1$ (Fig.1a), while the transitions  $1 \rightarrow 2$  and $0 \rightarrow 3$ are driven by two classical and constant fields with real Rabi frequencies $\Omega$ and $\Omega_0$, respectively. The classical fields create parametric coupling between two weak fields, while the $\Omega$ field provides EIT conditions for both quantum fields. We suppose $\Omega_{0} \gg \gamma_{3}$, with $\gamma_{3}$ the decay rate of the state $\mid 3 \rangle$, so that the bare atomic states $\mid 0 \rangle$ and $\mid 3 \rangle$ are split into a doublet of dressed states $\mid \pm \rangle=( |0 \rangle \pm | 3\rangle )/ \sqrt 2$ , which are well separated by 2$\Omega_{0}$ Fig.1b. If now the condition $\Delta= \omega_{10} - \omega_1 = \omega_{32}-\omega_2 = \Omega_0$ for the detunings of the quantum fields is fulfilled, where $\omega_1$ and $\omega_2$ are the carrier frequencies of the fields $\hat{\mathcal E}_1$ and $\hat{\mathcal E}_2$, respectively, then the quantum fields are in the exact resonance with the transitions $\mid + \rangle \rightarrow \mid 1 \rangle$ and $\mid + \rangle \rightarrow \mid 3 \rangle$ and their conversion into each other occurs with the same or higher efficiency as in the previous case of three-level atom \cite{gogmal}. Two-photon ladder type transition  in the Rb atom has been considered also in \cite{Chaneliere}, which involves three classical fields for two-photon excitation of the atom and creation of EIT conditions for signal and idler photons. So, this scheme is different from our system.
\begin{figure}[b]{\includegraphics*
[scale = 0.7]{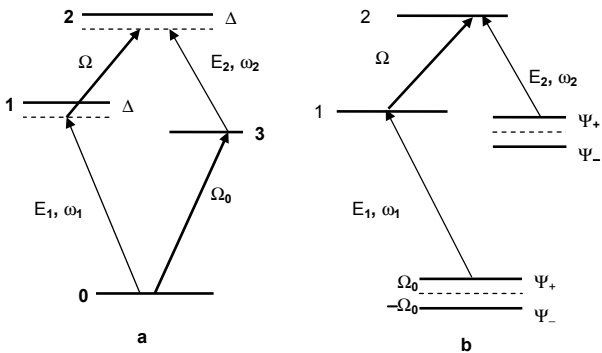}}
\caption{Scheme for conversion of quantum fields $E_{1}$ and $E_{2}$ with essentially different wavelengths in the basis of bare (a) and dressed (b) atomic states.}
\end{figure}

We consider the interaction of cold atoms with level configuration shown in Fig.1a with the two quantum fields
\begin{equation}
E_{1,2}(z,t)=\sqrt{\frac{\hbar \omega _{1,2}}{2{\varepsilon }_{0}V}}\hat{\mathcal E}_{1,2}(z,t)\exp [i(k_{1,2}z-\omega _{1,2}t)]+h.c.,
\end{equation}
co-propagating along the $z$ axis with $k_{1,2}$ the wave vectors of quantum fields, where $V$ is the quantization volume taken to be equal to interaction volume and the field operators $\hat{\mathcal E}_{i}(z,t)$ obey the commutation relations \cite{sis}
\begin{equation}\label{comrele}
\lbrack \hat{\mathcal E}_{i}(z,t),\hat{\mathcal
E}_{j}^{\dagger}(z,t^{\prime})]=\frac{L}{c}\delta _{ij}\delta
(t-t^{\prime}),
\end{equation}
where $L$ is the length of the medium. We describe the latter using atomic operators $\hat{\sigma}_{\alpha \beta} (z,t) = \frac{1}{N_{z}}\sum_{i=1}^{N_{z}}\mid \alpha \rangle _i \langle \beta \mid $ averaged over the volume containing many atoms $N_{z}=\frac{N}{L}dz\gg 1$ around position $z$, where $N$ is the total number of atoms. In RWA the Hamiltonian of the system in the interaction picture is given by
\begin{eqnarray}
H&=&\hbar \frac{N}{L}\int\limits_{0}^{L}dz \bigl (\Delta \hat{\sigma}_{11} +\Delta \hat{\sigma}_{22}-g_{1}\hat{\mathcal E}_{1}\hat{\sigma}_{10}e^{ik_{1}z}\nonumber \\
&-& g_{2}\hat{\mathcal E}_{2}\hat{\sigma}_{23}e^{ik_{2}z} -\Omega_0 \hat{\sigma_{30}} e^{ik_{0 }z}-\Omega \hat{\sigma_{21}}e^{ikz}+h.c. \bigr)
\label{ham}.
\end{eqnarray}
Here $k, k_{0}=\vec{k}\hat{e}_{z},\vec{k_0}\hat{e}_{z}$ are the projections of the wave-vectors of the driving fields on the $z$ axis and $g_{\alpha }=\mu _{\alpha \beta}\sqrt{\omega _{i}/(2\hbar {\varepsilon}_{0}V)}$ is the atom-field coupling constants with $\mu _{\alpha \beta }$ being the dipole matrix element on the atomic transition $\alpha \rightarrow \beta $. We consider the cold atomic ensemble for which the Doppler broadening is smaller than all relaxation rates and can be neglected. Using the slowly varying envelope approximation, the propagation equations for the quantum field operators take the form
\begin{eqnarray}\label{max1}
\left( \frac{\partial }{\partial z}+\frac{1}{c}\frac{\partial }{\partial t}\right) \hat{\mathcal E}_{1}(z,t)&=&ig_{1}\frac{N}{c}\hat{\sigma}_{01}e^{-ik_{1}z}+\hat{F}_{1}, \\
\left( \frac{\partial }{\partial z}+\frac{1}{c}\frac{\partial
}{\partial t}\right) \hat{\mathcal E}_{2}(z,t)&=&
ig_{2}\frac{N}{c}\hat{\sigma}_{32}e^{-ik_{2}z}+\hat{F}_{2}\label{max2}, \end{eqnarray}
where $\hat{F}_{i}(z,t)$ are the commutator preserving Langevin operators. We obtain the atomic coherences from the equation
\begin{equation}\label{atcoh}
\frac{d\hat{\sigma}_{ij}}{d t} = \frac{i}{\hbar} [H,
\hat{\sigma}_{ij}]+\frac{d\hat{\sigma}_{ij}}{d t} \bigl |_{rel},
\end{equation}
where the last term accounts for all the relaxations. We pass from the bare atomic states $\mid \phi \rangle = \{\mid 0 \rangle, \mid 1 \rangle, \mid 2 \rangle,\mid 3 \rangle\}$ to the dressed ones  $\mid \psi \rangle = S^\dagger U^ \dagger \phi = \{\mid - \rangle, \mid 1 \rangle, \mid 2\rangle,\mid + \rangle\}$ with
\begin{equation}
S =  \left( \begin{array}{cccc}
\cos\theta & 0 & 0 & -\sin\theta\\
0 & 1 & 0 & 0\\
0 & 0 & 1 & 0\\
\sin\theta & 0 & 0 & \cos\theta \end{array} \right), U= \left( \begin{array}{cccc}
1 & 0 & 0 & 0\\
0 & 1 & 0 & 0\\
0 & 0 & 1 & 0\\
0 & 0 & 0 & e^{ik_0 z} \end{array} \right),
\end{equation}
where $\tan 2\theta = 2\Omega_0/\Delta_0$. The transformation $U$ is applied in order all the atoms to experience the same $S$ transformation on the propagation line. In the dressed state and for $\Delta_0=\omega_{30}-\omega_0=0$ the Hamiltonian of the system in the dressed states basis$\tilde H =S^\dagger (U^\dagger H U) S$ takes the form
\begin{eqnarray}
\tilde{H}&=&\hbar \frac{N}{L}\int\limits_{0}^{L}dz \bigl [ \Delta \hat{\sigma}_{11} + \Delta \hat{\sigma}_{22} - \Omega_0 (\hat{\sigma}_{--}-\hat{\sigma}_{++})\nonumber \\
&-&\Omega \hat{\sigma}_{21}e^{ik z}-\frac{g_{1}\hat{\mathcal E}_{1}}{\sqrt{2}}e^{ik_{1}z} (\hat{\sigma}_{1-} - \hat{\sigma}_{1+})\nonumber \\&-&\frac{g_{2}\hat{\mathcal E}_{2}} {\sqrt{2}} e^{i(k_{2}+k_0)z} (\hat{\sigma}_{2-}+\hat{\sigma}_{2+})\bigr]+h.c. \end{eqnarray}
Then the atomic coherences $\hat \sigma_{01}$ and $\hat
\sigma_{32}$ are defined by $\hat \sigma_{\pm 1,2}$ as
\begin{eqnarray}
 \hat{\sigma}_{01} &=& ( \hat{\sigma}_{-1}- \hat{\sigma}_{+1}) /\sqrt2, \\
 \hat{\sigma}_{32}  &=& ( \hat{\sigma}_{-2}+ \hat{\sigma}_{+2}) /\sqrt2. \end{eqnarray}
In the weak-field (single-photon) limit, the equations for atomic coherences $\hat \sigma_{\pm 1} = \hat{\rho}_{\pm 1} e^{ik_{1}z}, \hat \sigma_{\pm 2}=\hat{\rho}_{\pm 2} e^{i(k_{2}+k_0)z} $ are treated
perturbatively in $\hat{\mathcal E}_{1,2}$. In the first order only $\langle \hat\sigma_{--}\rangle=\langle\hat \sigma_{++}\rangle=1/2$ are different from zero and we obtain:
\begin{eqnarray}
\dot{\hat{\rho}}_{-1,2}&=&-i(\Delta + \Omega_0-i\Gamma _{1,2+})\hat{\rho}_{-1,2}-\frac{\Gamma_3}{2}\hat{\rho}_{+1,2}\nonumber \\ &+&i\frac{g_{1,2}\hat{\mathcal E}_{1,2}}{\sqrt2}\hat \sigma_{--}+i\Omega \hat{\rho}_{-2,1}, \label{coh_m1}\\
\dot{\hat{\rho}}_{+1,2}&=&-i(\Delta - \Omega_0-i\Gamma _{1,2+})\hat{\rho}_{+1,2}-\frac{\Gamma_3}{2}\hat{\rho}_{-1,2}\nonumber \\ &\mp& i\frac{g_{1,2}\hat{\mathcal E}_{1,2}}{\sqrt2}\hat \sigma_{++}+i\Omega \hat{\rho}_{+2,1}, \label{coh_p2}
\end{eqnarray}
where $\Gamma_{1,2+}=\Gamma_{1,2}+\Gamma_3/2$, $\Gamma _{1,2}$ are the transverse relaxation rates, which in the case of cold atoms are simply $\Gamma _{1,2}=\gamma_{1,2}/2$, with $\gamma_{1,2}$ the natural decay rates of the excited states 1 and 2. Here the phase matching condition $\bigtriangleup k =k_1+k-k_2-k_0=0$ is supposed to be fulfilled.

It is seen from (\ref{coh_m1}), (\ref{coh_p2}) that for $\Delta=\Omega_0$ the polarizations $\hat \rho_{-1,2}$ are strongly suppressed compared to $\hat \rho_{+1,2}$ and can be neglected. The solution for $\hat \rho_{+1,2}$ are easily found from (\ref{coh_p2}) in the first order of $\hat {\mathcal E}_{1,2}$
\begin{eqnarray}
\hat{\rho}_{+1}&=& - i \frac{\Gamma_{2+}}{2\sqrt2 D^2} g_{1}\hat{\mathcal
E}_{1}-i\frac{\Omega ^{2}-\Gamma_{2+}^{2}}{2\sqrt2D^{4}} g_{1}\frac{\partial \hat{\mathcal E}_{1}}{\partial t}\nonumber \\& &- \frac{\Omega }{2\sqrt2 D^2}
g_{2}\hat{\mathcal E}_{2} +\frac{(\Gamma_{1+}+\Gamma_{2+}) \Omega }{2\sqrt2 D^{4}}g_{2}\frac{\partial \hat{\mathcal E}_{2}}{\partial t},\label{sol1}\\
\hat{\rho}_{+2} &=&+i \frac{\Gamma_{1+}}{2\sqrt2 D^2} g_{2}\hat{\mathcal
E}_{2}+i\frac{\Omega ^{2}-\Gamma_{1+}^{2}}{2\sqrt2D^{4}} g_{2}\frac{\partial \hat{\mathcal E}_{2}}{\partial t}\nonumber \\& & + \frac{\Omega }{2\sqrt2 D^2}g_{1} \hat{\mathcal E}_{1}-\frac{(\Gamma_{1+}+\Gamma_{2+}) \Omega }{2\sqrt2 D^{4}}g_{1}\frac{\partial \hat{\mathcal E}_{1}}{\partial t},\label{sol2}\\
D^2&=&\Omega^{2}+\Gamma_{1+}\Gamma_{2+}. \quad\quad\quad \quad\quad\quad \quad\quad\quad\nonumber
\end{eqnarray}
The first terms in right hand side (RHS) of Eqs.(\ref{sol1}), (\ref{sol2}) are responsible for linear absorption of
quantum fields and define the field absorption coefficients
$\kappa_{i}={g_{i}^{2}\Gamma N}/{4 c\Omega ^{2}}$ ($\Gamma \sim \Gamma_{1,2}$), upon substituting these expressions into Eqs.(\ref{max1}), (\ref{max2}). The second terms represent the dispersion contribution to the group velocities $v_i$ of the pulses leading to $v_{i}={4c\Omega ^{2}}/{g_{i}^{2}N}\ll c$, while the two rest terms describe the parametric interaction between the fields. However, we neglect the last terms in these equations, since they become strongly suppressed by the factor $\Omega ^{2}T/\Gamma \gg1$, with $T$ the initial pulse width.

For implementation of the QFC in a dense atomic medium three conditions must be fulfilled. The first, the photon absorption is negligibly small $\kappa_i L \ll 1$. The second, the initial spectrum of the quantum fields should be contained within the EIT window $\Delta \omega_{EIT} = {\Omega^{2}} / ({\Gamma \sqrt{\alpha }})$ \cite{fleis}, resulting in little pulse distortion from absorption, i.e. $\Delta \omega _{EIT}T \geq 1$, with $\alpha =\mathcal{N} \sigma L$ the optical depth, $\sigma =\frac{3}{4\pi }\lambda ^{2}$ the resonant absorption cross-section and $\mathcal{N}$ the atomic number density. And finally the pulse broadening should be minimal, showing that the spreading of the quantum pulses caused by the group-velocity dispersion is strongly reduced, which is achieved if $\frac{16L}{v_i T^{2}\Omega}\leq 1$.

Since in the absence of photon losses the noise operators
\begin{equation}\label{maxx1}
\left( \frac{\partial }{\partial
z}+\frac{1}{v_{j}}\frac{\partial}{\partial t}\right) \hat{\mathcal E}_{j}(z,t)=i\beta \hat{\mathcal E}_{k}\; , \; \; j,k=1,2,
\end{equation}
where $\beta ={g_{1}g_{2}N/4c\Omega }$ is the parametric coupling between the quantum fields.

For numerical estimations we choose a sample of $^{87}Rb$ vapor with the states $5S_{1/2}, 5P_{3/2},  4D_{3/2}$,  and $5P_{1/2}$ being the atomic states 0, 1, 2 and 3, respectively. In this case the quantum fields' wavelengths are $\lambda_{1}\sim 780nm, \lambda_{2}\sim$ 1,47$\mu m$ and $g_{2}/g_{1}\sim 0.96$ this provides the quantum fields $\hat{\mathcal E}_{1,2}$ to have almost equal group velocities on the corresponding transitions $v_1=v_2=v$, which leads to a simple solution of equations (\ref{maxx1}) in a form
\begin{equation}\label{field}
\hat{\mathcal E}_{i}(z,t)=\hat{\mathcal E}_{i}(0,\tau )\cos (\beta z)+i\hat{\mathcal E}_{j}(0,\tau )\sin(\beta z),
\end{equation}
where $i,j=1,2$ and $i \neq j$.

To show that the proposed scheme is suitable for converting
individual photons at one frequency to another frequency, while preserving initial quantum coherence of single-photon state, we analyze the evolution of the input state $\mid \psi_{in}\rangle = \mid 1_{1}\rangle \otimes \mid 0_{2}\rangle $ consisting of a single-photon wave packet at $\omega _{1}$ frequency, while
$\omega _{2}$ field is in the vacuum state. The similar results are clearly obtained in the case of one input photon at $\omega _{2}$ frequency. We assume that initially the $\omega _{1}$ pulse is localized around $z=0$ with a given temporal profile
$f_{1}(t)$
\begin{equation}\label{9}
\langle 0\mid \hat{\mathcal E}_{1}(0,t)\mid \psi _{in}\rangle =\langle 0\mid \hat{\mathcal E}_{1}(0,t)\mid 1_{1}\rangle =f_{1}(t).
\end{equation}
In free space, $\hat{\mathcal E}_{1}(z,t)=\hat{\mathcal
E}_{1}(0,t-z/c)$ and we have
\begin{equation}\label{10}
\langle 0\mid \hat{\mathcal E}_{1}(0,t-z/c)\mid 1_{1}\rangle
=f_{1}(t-z/c).
\end{equation}
The intensities of the fields at any distance in the region $0
\leq  z \leq  L$ are given by
\begin{equation}\label{11}
\langle I_{i}(z,t)\rangle =\mid \langle 0\mid \hat{\mathcal
E}_{i}(z,t)\mid \psi _{in}\rangle \mid ^{2}.
\end{equation}
The quantum efficiency $\eta$  of the process is determined as the ratio of the mean photon numbers $\eta=n_{2}(L)/n_{1}(0)$, where $n_{i}(z)=\langle \psi_{in}\mid \hat n_{i}(z)\mid \psi
_{in}\rangle$ with $\hat n_{i}(z)$ the dimensionless operators for number of photons that pass each point on the z axis in the whole time
\begin{equation}\label{12}
\hat n_{i}(z)=\frac{c}{L}\int dt\hat{\mathcal
E}_{i}^{\dagger}(z,t)\hat{\mathcal E}_{i}(z,t).
\end{equation}
Hence the quantum efficiency for input single-photon at frequency $\omega_1$ is easily found from (\ref{field})-(\ref{12}) to be \begin{equation}\label{qe}
\eta=\sin^2\beta L. \end{equation} The efficiency of the
process reaches $100\%$ when $\beta L = \pi/2$. This is the case, e.g. if we take $L=1.6 mm, \Omega = 8\Gamma, \Gamma = 2\pi \times 3 MHz, \mathcal N =10^{13} cm^{-3}$. It can be easily checked that all conditions for efficient FC mentioned above are fulfilled with $T\sim 20ns$. The Doppler broadening can be neglected, if $ku \sim \Gamma$, where $u$ is the mean thermal velocity of the Rb atoms. From this condition the temperature of Rb vapor is obtained to be $T_{at}\sim0.03K$. These parameters appear to be within experimental reach, including the initial single-photon wave packets with a pulse length of several tens of nanoseconds \cite{mats, yuan} and the Rb vapor temperature with the mentioned densities \cite{Ketterle, Ketterle1}. From (\ref{maxx1}) and (\ref{qe}) it is evident that the shape of the photon pulse is conserved during the FC process.

A very important property of the scheme is the preserving of the initial quantum state of the weak field during the conversion. To show this we describe the output wave-packets of different frequencies by the wave functions $\Phi_{i}(L,t)= \langle 0\mid \hat{\mathcal E}_{i}(L,t)\mid \psi_{in}\rangle$ and introduce the creation operators $c^{\dagger}_i$ of the wave-packets associated with these mode functions as
\begin{equation}
\hat c_{1,2}^{\dagger}=N_{i}^{1/2}\int dt \Phi_{1,2}(L,t)\hat{\mathcal
E}_{1,2}^{\dagger}(L,t),
\end{equation}
with the normalization
constant $N_{i}=\frac {c}{L}(\int dt \mid \Phi_{1}(L,t) \mid^2)^{-1}$. These operators create the single-particle states in the usual way by acting on the vacuum state $\mid 0\rangle$
\begin{equation}
\hat c_{i}^{\dagger}\mid 0\rangle=\mid 1_{i}\rangle \end{equation}
and have the standard boson commutation relations
\begin{equation}\label{comrelc}
[\hat c_{i},\hat c_{j}^{\dagger}]=\delta_{ij}, \end{equation}
following from (\ref{comrele}). Correspondingly, the mean photon number at each mode is given by $n_{i}(L) = \frac{c}{L} \int dt \mid \Phi_{i}(L,t)\mid^2$. Then we obtain the output single-photon state as the eigenstate of total photon number operator, using (\ref{comrele}) and (\ref{12})
\begin{equation}
(\hat n_{1}(L)+\hat n_{2}(L))\mid \psi_{out}\rangle=\mid
\psi_{out}\rangle,
\end{equation}
which yields
\begin{equation}
\mid \psi_{out}\rangle=\frac{c}{L}\int dt
[\Phi_{1}(L,t)\hat{\mathcal
E}_{1}^{\dagger}(L,t)+\Phi_{2}(L,t)\hat{\mathcal E}_{2}^{\dagger}(L,t)]\mid
0\rangle.
\end{equation}

Note that this definition of quantized wave packets is only
useful, if the mode spectra are much narrower compared to the mode spacing that has been suggested from the very beginning. Now, for the algebra (\ref{comrelc}) we choose the representation of infinite product of all vacua
\begin{equation}
\mid 0\rangle=\prod_{i}\mid 0_{i}\rangle=\mid 0_{1}\rangle\mid
0_{2}\rangle\prod_{i\neq 1,2} {\mid 0_{i}\rangle}.
\end{equation}
However, since in our problem we deal with two frequency modes, while the other modes are not occupied by the photons and,  hence, are not taken into account during the measurements, the vacuum may be reduced to $\mid 0\rangle=\mid 0_{1}\rangle\mid 0_{2}\rangle$. For the sake of simplicity we consider an input single photon entangled in two well-separated temporal modes or time bins \cite {brend}
\begin{equation}
\mid \psi_{in} \rangle = (a \mid 1_{1}\rangle_{t} \mid
0_{1}\rangle_{t+\tau} +  b \mid 0_{1}\rangle_{t} \mid
1_{1}\rangle_{t+\tau})\otimes \mid 0_{2}\rangle, \end{equation}
where $\mid 0_{1}\rangle_{t}$ and $\mid 1_{1}\rangle_{t}$ denote Fock states with zero and one $\omega_{1}$ photon, respectively, at the time $t$ and $\mid a\mid ^{2} +\mid b \mid^{2}=1$, $\tau$ being the time shift between the temporal modes. Suppose that the single-photon wave packets $\mid 1_{1}\rangle_{t}$ and $\mid 1_{1}\rangle_{t+\tau}$ are characterized by temporal profiles $f_{0}(t)$ and $f_{\tau}(t)$, respectively, which are not overlapped in time due to $\tau\gg T$. Then, using (\ref{field}) and (\ref{10}), the wave function of output $\omega_{2}$ mode is readily calculated to be
\begin{equation}
\Phi_{2}(L,t)= \langle 0\mid \hat{\mathcal E}_{2}(L,t)\mid
\psi_{in}\rangle=a\Phi_{2,0}(L,t)+b\Phi_{2,\tau}(L,t),
\end{equation}
where
\begin{equation}
\Phi_{2,0(\tau)}(L,t)=-i\beta \int\limits_{0}^{L}dxf_{0(\tau)}(t)
J_{0}(\psi ).
\end{equation} Consequently, from (22) the
creation operator $c_{2}^{\dagger}$ can be represented as a sum of creation operators of the two temporal modes at $\omega_{2}$ frequency. Following the procedure discussed above the output
state in the case of complete conversion ($r_{1}=0$) is  eventually found in the form
\begin{equation}
\mid \psi_{out} \rangle = \mid 0_{1}\rangle\otimes(a \mid
1_{2}\rangle_{t} \mid 0_{2}\rangle_{t+\tau} +  b \mid
0_{2}\rangle_{t} \mid 1_{2}\rangle_{t+\tau}),
\end{equation}
showing that the initial $\omega_{1}$ qubit is transformed into another at $\omega_{2}$ frequency with the same complex amplitudes $a$ and $b$, thus preserving the original amount of entanglement, which is possible to verify experimentally.

In summary, the present work shows that the conversion of narrow-band IR and visible photons is noiseless and can be realized with a maximum quantum efficiency.  In addition, an initial quantum information content is fully preserved during the QFC, that may find important applications for interfering the telecommunication lines with atomic memory elements in quantum networks.

The author is thankful to S. Gu\'erin  and Yu. Malakyan for useful and simulating discussions. This work was supported by the Armenian Science Ministry Grant No 096.

\end{document}